\documentclass[epsf,usegraphicx]{mn2e}

\title[Spin up in  RX J0806+15]{Spin up in RX J0806+15 - the shortest period binary }

\author[Hakala et al]
{Pasi Hakala$^{1}$, Gavin Ramsay$^{2}$, Kinwah Wu$^{2}$,
 Linnea Hjalmarsdotter$^{1}$, \and Silva J\"arvinen$^{3,4}$, 
Arto J\"arvinen$^{3,4}$, Mark Cropper$^{2}$\\  
$^{1}$Observatory, University of Helsinki, PO Box 14, 
FIN-00014 University of Helsinki, Finland\\ 
$^{2}$Mullard Space Science Laboratory, University College London,
Holmbury St.\ Mary, Dorking, Surrey, RH5 6NT, UK\\
$^{3}$Nordic Optical Telescope, Apartado 474, E-38700 Santa Cruz de La Palma,  Canarias, Spain \\
$^{4}$Astronomy Division, PO Box 3000, 90014 University of Oulu, Finland \\
}
\date{Received: -- Accepted: --}

\begin{document}
\outer\def\gtae {$\buildrel {\lower3pt\hbox{$>$}} \over 
{\lower2pt\hbox{$\sim$}} $}
\outer\def\ltae {$\buildrel {\lower3pt\hbox{$<$}} \over 
{\lower2pt\hbox{$\sim$}} $}
\newcommand{\ergscm} {ergs s$^{-1}$ cm$^{-2}$}
\newcommand{\ergss} {ergs s$^{-1}$}
\newcommand{\ergsd} {ergs s$^{-1}$ $d^{2}_{100}$}
\newcommand{\pcmsq} {cm$^{-2}$}
\newcommand{\ros} {\sl ROSAT}
\newcommand{\exo} {\sl EXOSAT}
\def\rchi{{${\chi}_{\nu}^{2}$}}
\newcommand{\Msun} {$M_{\odot}$}
\newcommand{\Mwd} {$M_{wd}$}
\def\Mdot{\hbox{$\dot M$}}
\def\mdot{\hbox{$\dot m$}}

\maketitle

\begin{abstract}

RX J0806+15 has recently been identified as the binary system with the
shortest known orbital period. We present a series of observations of
RX J0806+15 including new optical observations taken one month apart.
Using these observations and archival data we find that the period of
this system is decreasing over time. Our measurements imply $\dot{f} =
6.11 \times 10^{-16} $ Hz/s, which is in agreement with a rate expected
from the gravitational radiation for two white dwarfs orbiting at a
given period. However, a smaller value of $\dot{f} = 3.14 \times
10^{-16} $ Hz/s cannot be ruled out. Our result supports the idea that
the 321.5 s period is the orbital period and that the system is the
shortest period binary known so far and that it is one of the
strongest sources of constant gravitational radiation in the sky. 
Furthermore, the decrease of the period strongly favours 
the unipolar inductor (or electric star) model 
rather than the accretion models.

\end{abstract}

\section{Introduction}

For accreting stellar binary systems in which the secondary star (mass
donating) is on the main sequence, the minimum orbital period is
$\sim$80 mins (Paczynski \& Sienkiewicz 1981). For secondary stars not
on the main sequence (such as helium stars), systems can also have 
orbital periods less than 80 min (Savonije, de Kool \& Van den Heuvel 1986 and references
therein). The other option for short period systems is that
the secondary star must be degenerate forming a double degenerate binary with a compact primary. 
A number of these systems have been known for some time although it is only
relatively recently that their nature has been revealed (cf Warner
1995). The orbital periods of these systems lie in the tens of mins
range.

Recently a group of three binaries have been discovered in which the
orbital period is thought to be less than $\sim$10mins: ES Cet
(KUV01584$-$0939) 10.3 min (Warner \& Woudt 2002), V407 Vul (RX
J1914+24) 9.5 mins (Cropper et al 1998) and RX J0806+15, 5.4min
(Ramsay, Hakala \& Cropper 2002a, Israel et al.\ 2002). If these periods
have been correctly identified as the binary orbital period, then these
have the shortest known binary periods. As
such they should be strong sources of gravitational waves.

In the case of ES Cet, strong He lines are seen in optical spectra
(Wegner et al.\ 1987) which have recently been found to be double peaked
(P.\ Woudt, priv.\ comm.) suggesting the presence of a disk. In the case of
V407 Vul, no emission lines are seen (Ramsay et al.\ 2002b) while RX
J0806+15 shows only weak He lines (Israel et al.\ 2002). Very recently 
(Israel et al.\ 2003) demonstrated that the optical and X-ray emission 
of RX J0806+15 are in anti-phase, as in the case of V407 Vul (Ramsay et al.\ 2000).

Currently there are several models to explain the observations of V407
Vul and RX J0806+15. A neutron star primary and an intermediate polar
model cannot be excluded although they are considered unlikely. The most
likely models all have a white dwarf primary and secondary. 
Two models (a polar and a direct accretor model) are
driven by mass transfer while one (a unipolar inductor or electric
star model) is driven by an electrical current. Of these models the
unipolar inductor model fits the observational properties better than
the accretion models, although these cannot be ruled out 
(Cropper et al.\ 2003 (review), Wu et al.\ 2002, Marsh \& Steeghs 2002, 
Ramsay et al.\ 2002a and Norton et al.\ 2002).

The best method to differentiate between these models is to
measure the period change of these systems. For systems driven by mass
transfer the period should increase over time 
(eg Savonije, de Kool \& Van den Heuvel 1986,
Nelemans et al.\  2001). In the case of
the unipolar inductor model, the system should be driven entirely by
gravitational radiation and the period should decrease (Wu et al.\ 2002).

Strohmayer (2002) showed using {\sl ROSAT} observations that for V407
Vul the period is {\sl decreasing} at a rate consistent with that
predicted from the emission of gravitational radiation.  In this paper
we present observations of the 5 min system RX J0806+15 and determine
the change in the period. We then discuss the implications of this
result.

\section{Observations}

In order to measure possible period changes in RXJ0806+15, we have
obtained data from the Nordic Optical Telescope La Palma 
between 5th and 8th of January 2003 and 5th and 7th of February
2003. These observations consist of high speed CCD photometry with an
approximate time resolution of 15 sec (10 sec integration time).  The
observations were carried out using ALFOSC in imaging mode without any
filter. We binned the CCD chip by 
a factor of two in both directions
and selected only a small 100x100 subwindow for fast readout. A total
of 4687 images were obtained. These were bias corrected and
flatfielded in the usual manner.  Conditions in January 2003 were
photometric and the seeing was typically better than 1". In February,
however, seeing was poorer (about 1".5).

Additionally some archival ESO VLT data from 2001 November 12 was analysed
to provide further constraints for a period from our earlier (January
2002) NOT run. There were a total of 214 data points that were
combined with our 1314 datapoints from two nights in January
2002. These data were also reduced using standard procedures. All of the
data were heliocentric corrected.

There are also some archival {\sl Chandra} observations available. However,
given the length of the observation (20 ksec), we have not included it here.  

\section{Period analysis}

We have three different datasets that can be used to search for the
change in the orbital period (Table 1). Chronologically these are: I) The {\sl ROSAT}
observations from 1994-95, II) The combined VLT data from Nov 2001 and
NOT data from Jan 2002 and finally III) The NOT data from Jan-Feb 2003.
We will now proceed to estimate the best period for each of these datasets
and then use the results to derive the change in the orbital period.

\begin{figure} 
\includegraphics[scale=0.5]{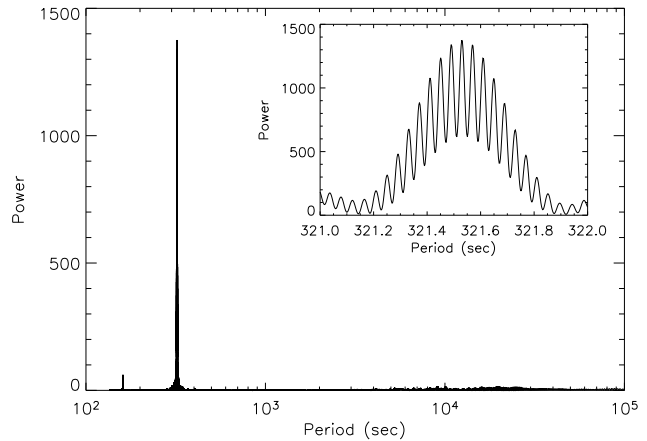}
 \caption{The Lomb-Scargle power spectrum of the Jan-Feb 2003 data and (inset) 
the same plot zoomed in near the best period.}
\end{figure}

The earliest of our three datasets is the 1994-1995 {\sl ROSAT} data
(Burwitz \& Reinsch 2001).  They found two possible values for the
best period in their {\sl ROSAT} data, namely: 321.5393 or 321.5465 s ($\pm$
0.0004 s). Their preference was the first of these. For consistency we
have re-analysed the {\sl ROSAT} data and confirm their results.

For our optical datasets we have used the Lomb-Scargle power spectrum
(Scargle 1982) to search for the periods. However, in order to obtain reliable
error estimates (crucial for the $\dot{P}$ measurement) we have
performed Monte Carlo simulations of our time series. This enables us
to take into account the true effects of aliasing in a rigorous
manner. Our Monte Carlo approach consists of following steps. i) Find
the best period using the Lomb-Scargle algorithm. ii) Fix the period
and fit the light curve shape with a 2nd order Fourier fit. iii) Then use
the (Gaussian) errors from the Fourier fit to produce 10000
synthetic datasets using the same time points as in the original
dataset and add noise comparable to the noise in real data. iv)
Finally analyze the 10000 datasets around the best period. To do this
in a reasonable amount of CPU time, we have used the minimum string
length method (Dwortesky, 1983) which can be easily restricted to the immediate
neighbourhood of the best period.

\begin{table*}
\begin{tabular}{llrr}
\hline 
Observation & Dates (HJD-2400000.0) & Best period (s) & Error (s)\\ 
\hline
{\sl ROSAT} (1994 \&1995) &  49648.9-49658.2 & 321.5393 OR 321.5465 & 0.00040 \\ 
                    &  49822.9-49827.0 &                      & \\
VLT (Nov 2001) \& NOT (Jan 2002) & 52226.258-.373 & 321.53352 & 0.00034 \\
					          & 52289.533-.653 & & \\
                                                 &  52290.354-.740 & & \\ 
NOT (Jan -Feb 2003) & 52645.443-.674 & 321.52957 & 0.00045  \\
                                   &   52646.453-.639 & & \\
                                    &  52647.435-.773 & & \\
                                     &  52676.420-.438 & & \\
                                    &   52677.493-.576 & & \\
\hline
\end{tabular}
\caption{The times of observation and best periods for our three 
different datasets}
\label{periods}
\end{table*}

\begin{figure}
\includegraphics[scale=0.5]{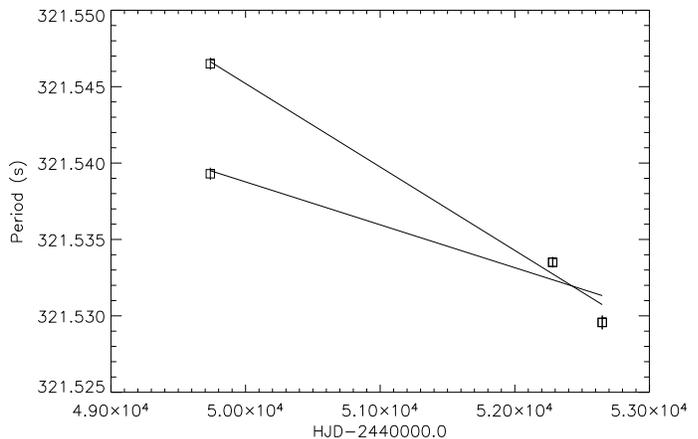}
 \caption{The fits to the change in period assuming two different values for the {\sl ROSAT} period.}
\end{figure}

\begin{figure}
\includegraphics[scale=0.5]{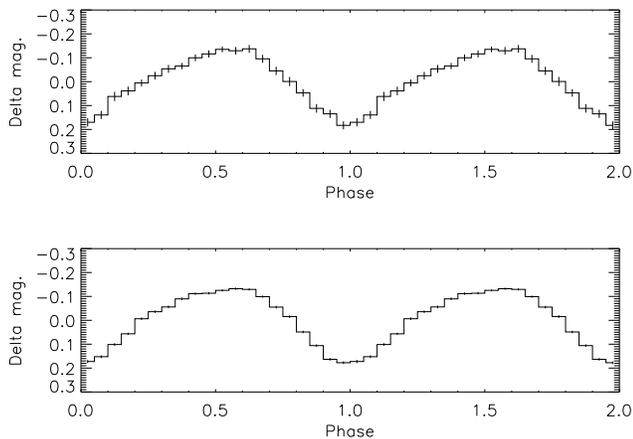} \caption{The January 2002 (top) and January-February
 2003 white light light curves folded on the orbital period.}
\end{figure}

Our second dataset, which consists of VLT data from Nov 2001 and NOT
data from Jan 2002 is problematic in terms of aliasing. Both of these
runs are too short to 'phase' them together in a unique manner and as
a result severe aliasing arises . However, we can use some {\it a
priori} information from our later, Jan-Feb 2003, NOT run to remove
the aliasing (i.e. pick the right spike from a handful of possible spikes
suggested by the Lomb-Scargle analysis and the subsequent 
Monte Carlo simulation).

The $\dot{P}$ from Jan 2003 data combined with the {\sl ROSAT} runs is about
1.2 msec/year. Further, our best period from NOT 2003 run is
321.52957s $\pm$ 0.00045s (see below). Assuming these values we can estimate
that the period in Jan 2002 should be approximately  321.532s. The possible spikes present
in the periodogram are 321.515s, 321.534s and 321.552s.
 Only one of these (321.534s) has a period close to the {\sl ROSAT} and NOT 2003 periods.
  If we now assume this is the 'correct' power peak,
 we can use our Monte Carlo study to get 321.53352s $\pm$ 0.00034s for our second run.

The third run consists of data from three consecutive nights in Jan
2003 (5th-8th) and from two nights in Feb 2003 (5th and 6th). This run
does not suffer from serious aliasing and using the Monte Carlo
technique outline above, we find 321.52957 $\pm$ 0.00045 s as the best
period.  We have plotted the periodogram of the NOT 2003 run in Figure
1. The inset shows the part near the best period.

\section{The period change}

To measure the possible change in the orbital period. We have fitted
our periods from three different epochs (defined as the mean of the
time points from each of the epochs) with a linear fit (Figure 2). 
Depending on the choice of which {\sl ROSAT} period we use, 
we obtain two sets of frequency ($f$) 
and $\dot{f}\ (\equiv - \dot{P}/{P^{2}})$.    

If we adopt 321.5393s as the true {\sl ROSAT} period (as
preferred by Burwitz \& Reinsch, 2001).  We get $\dot{f}$ =
$3.14\times10^{-16}$ Hz/s. On the other hand, the longer {\sl ROSAT} period
321.5465 s yields $\dot{f} = 6.11\times10^{-16}$ Hz/s. In both cases
the error is $2.0\times10^{-17}$ Hz/s. Judging from our fit, the latter period
of the two is better. 
Neither of the fits is very good, but  given the number of measurements
to be fitted (three),  we cannot make a definitive conclusion between the two
solutions.

We have also folded both the 2002 and 2003 NOT data over the orbital period and 
plotted that in Figure 3. In order to do this, we need to adopt a value
for $\dot{f}$ that gives us the correct phasing. This can be used as 
a feasibility test for our model where the period is decreasing.
We find two epochs when the flux should be at minimum, one just
before the 2002 NOT dataset and another before the 2003 NOT dataset.
These are HJD: 2452288.999256 and 2452645.436696. Now, to produce
correct phasing an integer number of (changing) periods must have passed
in between these two epochs. As we expect the period change to be linear,
at least in short time scale, we can estimate an average number of periods
that have passed (95780) using the average of the best periods from the 2002
and 2003 datasets as the mean period. We can then check if we can 
produce the  correct phasing using our best periods for the two epochs 
together with the two possible values of $\dot{f}$, and their errors. 
As a result we conclude that the data can be phased together using
either of the two $\dot{f}$ values. The larger $\dot{f}$ produces a better
fit, but we cannot exclude the smaller $\dot{f}$. 

 The resulting phase folded light curves are plotted in Figure 3. The modulation 
and its asymmetry is roughly the same in both datasets. 
However, it seems like the 2003 data 
shows a broader, flatter flux maximum than the earlier dataset. The two datasets 
are identical in terms of instrumentation used.

\section{Discussion}

The current competing models for the system can be classified into two
types: accreting and non-accreting models (cf \S 1). The accreting and
the non-accreting models  
can be distinguished by determining the period change in the system.

As $\dot f/f = -\dot P/P$, 
  we can calculate $\dot f$ from the orbital parameters of the system.  
For an accreting system, the period change is given by:
\begin{eqnarray}
  {{\dot P} \over P }
     & = & 3 {{{\dot J}_{\rm o}}\over {J_{\rm o}} }
       -  3 {{{\dot M}_2}\over{M_2}} (1 - q)    \nonumber
\end{eqnarray}
\noindent 
(Appendix A), where $J_{\rm o}$ is the orbital angular
momentum, $M_2$ is the mass of the secondary star, $q$ is the ratio of
the mass of the secondary star to the primary star.      
For systems with matter transfer from the
secondary to the primary, $\dot M_2$ is negative
whereas $(1-q)$ is positive. In general, ${{\dot P}/P}$ may take
a positive or negative value depending on $J_{\rm o}$ and the driving
mechanism of the mass transfer. However, if the secondary is a
degenerate star, ${{\dot P}/ P}$ is positive 
  (i.e.\ negative ${{\dot f}/ f}$) 
  when the angular momentum
loss from the system is due to gravitational radiation (see Ritter 1986).
  
For non-accreting system with {\sl no} coupling 
  between the stellar spins and the orbit, 
\begin{eqnarray}
    {{\dot P} \over P}  & = &
    - {96 \over 5} \left( {{GM}\over {c^2 a} } \right)^3
      \left({c \over a}\right)
      \left[{q\over {(1+q)^2}} \right]    \ ,  \nonumber
\end{eqnarray}
\noindent
   where $G$ is the gravitational constant,
   $c$ is the speed of light
  $M$ is the total mass of the two stars,
   and $a$ is the orbital separation.
If the spin-orbit coupling occurs and 
  the the angular momenta of the orbit and stellar spins 
  maintain roughly constant ratios, then 
\small
\begin{eqnarray}
   {{\dot P} \over P}
    & = & - {96 \over 5} \left( {{GM}\over {c^2 a} } \right)^3
      \left({c \over a}\right)
      \left[{q\over {(1+q)^2}} \right]      \nonumber  \\
     &  &  \hspace*{-0.75cm} \times \left\{ \ 1 - {6 \over 5}
     \left[
     \alpha_1 \left( {{1+q} \over q}\right) \left({{R_1}\over a}\right)^2
     + \alpha_2 \left({1+q}\right)\left({{R_2}\over a}\right)^2 \right] \right\}^{-1}
         \nonumber
\end{eqnarray}
\normalsize
\noindent 
(Appendix A), 
   where $R_1$ and $R_2$ are the radii of the primary and secondary
   respectively, $\alpha_1$ and $\alpha_2$ are the coupling parameters
   of the spins of the two stars with the orbit.    
For parameters typical of double degenerated systems, 
${{\dot P}/ P}$ is negative 
and hence ${{\dot f}/ f}$ is positive 
(with and without spin-orbit coupling), 
which is in contrast to the prediction by the accreting models.
 
If the 321.5 s period is the orbital period of the system, 
the observed period change is not consistent
with the prediction of the accreting models but is consistent with
non-accreting models. If we interpret that the observed period is the
spin period of the white dwarf in an intermediate polar, both spin up
and down is possible. However, we consider that the
spin-interpretation is unlikely (Cropper et al.\ 2003).

In contrast, the detected decrease in the orbital period strongly
favours the non-accreting models, in particular the unipolar-inductor
model (Wu et al.\ 2002), in which the two stars in the systems are white
dwarfs. In this model, angular momentum loss from the system and
system evolution is due to gravitational radiation and the spin-orbit
coupling process (if present) via unipolar induction.
  
In the abscence of accretion the secondary must reside well within its
Roche lobe.  From the Roche-lobe relation of Eggleton (1983) and the
Nauenberg (1972) mass-radius relation for white dwarfs, we find that
for $M_{1}=1.0~M_\odot$ and an orbital period of 321.5~s, the white
dwarf secondary, $M_{2} > 0.14~M_\odot$ in order that it stays within
its Roche lobe.

Now, suppose that the observed period is the orbital period and that
the change is due to orbital evolution.  We show in Figure 4 the
predicted change in frequency for a range of component masses, for the
case with no spin-orbit coupling.  Using the longer {\sl ROSAT}
period, we find that for $M_{1}$=1.0~M$_\odot$ a secondary mass of 0.3~M$_\odot$
is required.  For a less massive primary, a more massive secondary is required. 
 Taking the shorter {\sl ROSAT} period, a $M_{1}$=1.0~M$_\odot$ primary requires 
a secondary of mass $>$0.15~M$_\odot$. The studies of double white dwarf
systems (eg Marsh 2000, Napiwotzki et al.\ 2002) have shown that all such systems
tend to have a mass ratios of the order of 1.0. Furthermore, the individual white
dwarf masses in these systems are below 0.5 M$_\odot$. Such masses are compatible
with the measured period change. In all of the above mentioned
cases, the deduced masses for the secondary are consistent with
requirement that the star is detached from its critical Roche surface.

If the masses of the system could be determined independently,
the degree to which the period decrease differs from that expected from
gravitational radiation losses alone would provide information on the
degree of asynchronism in the spin of the primary. This asynchronism
is the origin of the power generation in the unipolar inductor model, so continuing
timing observations are essential to determine whether this is
consistent with the observed luminosity and reasonable component masses.

To conclude we note that for perfectly acceptable masses of the components, 
the $\dot{f}$ is compatible with gravitational radiation. This is consistent with the
results for V407 Vul (Strohmeyer 2002).
The implication is that there is no mass transfer in either of these systems, 
favouring the unipolar inductor (electric Star) model.

\begin{figure}
\includegraphics[scale=0.5]{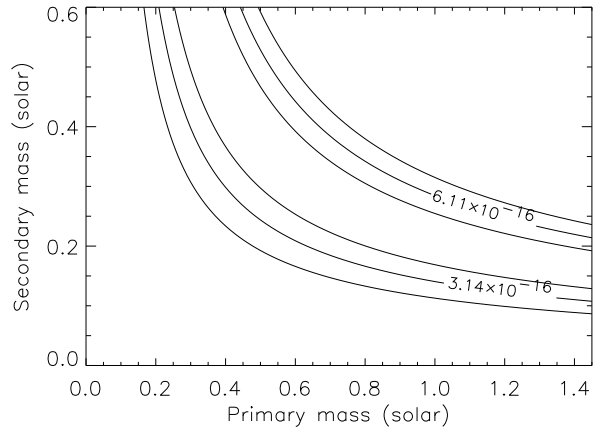} 
\caption{The $\dot{f}$ as a
 function of the component masses (assuming no spin-orbit coupling). 
 The two labelled contours show the
 measured $\dot{f}$ for the two different {\sl ROSAT} periods and the
 contours next to them give the 3 sigma limits.}
\end{figure}

\section{Acknowledgements}

PJH is an Academy of Finland research fellow.  Based on observations made with 
the Nordic Optical Telescope, operated jointly by Denmark, Finland,
Iceland,  Norway, and Sweden in La Palma. 
The data presented here have been taken using ALFOSC, which is owned 
by the Instituto de Astrofisica de Andalucia (IAA) and operated 
at the NOT.  We would like to thank 
the referee Dr G. Nelemans for very useful suggestions.
{}
\appendix 
\section{Orbital evolution} 

Consider a binary with masses  $M_1$ and $M_2$ and
  radii $R_1$ and $R_2$.  
Let the binary orbital period be $P$ and the orbital separation be $a$.  
Then  $\omega\ (\equiv 2\pi/P)$ is orbital angular velocity, and
  $M$ is the total mass.  The angular momenta of the
  orbital, star 1 and star 2 are respectively
\begin{eqnarray}   
   J_{\rm o}  & = & {{M_1M_2}\over {M}}a^2\ \omega \ , \nonumber \\ 
   J_1 & = & {2\over 5} M_1 R_1^2\ \omega_1 \ , \nonumber \\  
   J_2 & = & {2\over 5} M_2 R_2^2\ \omega_2 \ ,  \nonumber  
\end{eqnarray}  
   where $\omega_1$ and $\omega_2$ are the spin angular velocity of the two stars
(Assuming a spherical secondary). 

Define two spin parameters  
  $\alpha_1 \equiv \omega_1/\omega$  and  $\alpha_2 \equiv \omega_2/\omega$. 
The total angular momentum of the binary system can then be expressed as  
\begin{eqnarray}   
   J & = & J_{\rm o} + J_1 + J_2 \nonumber \\ 
     & = & \lambda_{\rm o}\ \omega^{-1/3} 
        + \lambda_*\ \omega \ ,  \nonumber 
\end{eqnarray}    
   where the two coefficients $\lambda_{\rm o}$ and $\lambda_*$ are 
\begin{eqnarray}  
   \lambda_{\rm o} 
       & = & {{G^{2/3}M_1M_2}\over{(M_1+M_2)^{1/3}}}  \ ,  \nonumber \\
   \lambda_*  & = &   
      {2\over 5}\left(\alpha_1 M_1 R_1^2 + \alpha_2 M_2 R_2^2  \right) \ .  \nonumber 
\end{eqnarray} 
If there is no mass exchange between the two stars,
 $d \lambda_{\rm o}/dt =0$.
Moreover, if the spin-orbit coupling (via tidal or via magnetic field)
 maintains roughly constant angular momentum ratios 
 between the orbit and the stars, we may set $d\lambda_*/dt = 0$
 and obtain a simple relation that relates the time derivatives of $J$ and $\omega$:
\begin{eqnarray}  
   \dot J & = & -{1\over 3 } \left({{\lambda_{\rm o}}\over { \omega^{1/3}}}\right)
      \left[~1 - 
       3 \left({{\lambda_*}\over{\lambda_{\rm o}}}\right) \ \omega^{4/3} \right] 
      \ {{\dot \omega}\over {\omega}}  \ . \nonumber 
\end{eqnarray}    

For systems with angular-momentum loss due to gravitational radiation, 
\begin{eqnarray}  
   \dot J & = &  - {32 \over 5} {G \over c^5} 
    \left( {{M_1^2 M_2^2} \over {M^2}}  \right) 
       a^4 \omega^5   \nonumber 
\end{eqnarray}     
 (see e.g.\ Landau and Lifshitz 1958), where $c$ is the speed of light. 
It follows that   
\small
\begin{eqnarray}
   {{\dot P} \over P}
    & = & - {{\dot \omega}\over {\omega}} \nonumber \\
    & = & - {96 \over 5} \left( {{GM}\over {c^2 a} } \right)^3
      \left({c \over a}\right)
      \left[{q\over {(1+q)^2}} \right]      \nonumber  \\
     &  &  \hspace*{-0.70cm} \times 
     \left\{ \ 1 - {6 \over 5}
     \left[
     \alpha_1 \left( {{1+q} \over q}\right) \left({{R_1}\over a}\right)^2     
     + \alpha_2 \left({1+q}\right)\left({{R_2}\over a}\right)^2 \right] \right\}^{-1}
      \ .  \nonumber
\end{eqnarray}
\normalsize
The first term above is the ratio of the gravitation radius of the binary 
  to the orbital separation; 
  the second term is the reciprocal of the light crossing time of the binary; 
  the square bracket term can be considered as a structural factor of the system; 
  and the last term is the spin-orbit coupling  correction . 
For systems with the spins of the stars decoupled from the orbit, 
  $\alpha_1 = \alpha_2 = 0$, and  
\begin{eqnarray}
    {{\dot P} \over P}  & = &
    - {96 \over 5} \left( {{GM}\over {c^2 a} } \right)^3
      \left({c \over a}\right)
      \left[{q\over {(1+q)^2}} \right]    \ . \nonumber
\end{eqnarray}

From above, we can also see that 
  ${\dot P}/P$ is always negative when the $\{....\}$ term is positive. 
This can be satisfied easily 
  for positive $\alpha_1$ and $\alpha_2$.  
This implies that the orbit period of a binary system decreases 
  because the gravitational radiation that it emits 
  carries aways the orbital angular momentum, 
  and if in the presence of spin-orbit coupling 
  orbital angular momentum is also extracted  
  to synchronise the spins of the stars. 
  We note that for systems with zero mass loss 
  but non-zero mass exchange between the two stars,  
\begin{eqnarray}  
   {{{\dot J}_{\rm o}}\over {J_{\rm o}} } & = & 
        {{{\dot M}_1}\over{M_1}} + {{{\dot M}_2}\over{M_2}}   
          + {1\over 2 }{{\dot a }\over a } \ \nonumber  \\ 
        & = & {{{\dot M}_2}\over{M_2}} (1 - q)  
          - {1\over 3 }{{\dot \omega}\over \omega }  
        \  . \nonumber 
\end{eqnarray}   
It follows that  
\begin{eqnarray}
  {1 \over 3}  {{\dot P} \over P }
     & = & {{{\dot J}_{\rm o}}\over {J_{\rm o}} }
       -  {{{\dot M}_2}\over{M_2}} (1 - q)  \ . \nonumber
\end{eqnarray}
(Here, ${\dot J}_{\rm o}$ 
   consists of contributions of orbital angular momentum loss 
   by gravitational radiation and by spin-orbital coupling.)

\end{document}